\documentclass[aps,prb,twocolumn,epsfig,floatfix,twoside,a4paper,showpacs]{revtex4}
\usepackage{epsfig}

\begin{document}
\input epsf
\title{Thermodynamic behaviour and structural properties of an 
aqueous sodium chloride solution upon supercooling}
\author{
D.~Corradini,
P.~Gallo\footnote[1]{Author to whom correspondence should be addressed; 
e-mail: gallop@fis.uniroma3.it} and M.~Rovere }
\affiliation{Dipartimento di Fisica, Universit\`a ``Roma Tre''\\
and Democritos National Simulation Center, \\ 
Via della Vasca Navale 84, I-00146 Roma, Italy\\}

\begin{abstract}
\noindent
We present the results of a molecular dynamics simulation study of 
thermodynamic and structural properties upon supercooling of a low concentration 
sodium chloride solution in TIP4P water and the comparison with the 
corresponding bulk quantities. 
We study the isotherms and the isochores for both the aqueous solution and
bulk water. The comparison of the phase diagrams shows that thermodynamic properties 
of the solution are not merely shifted with respect to the bulk. 
Moreover, from the analysis of the thermodynamic
curves, both the spinodal line and 
the temperatures of maximum density curve 
can be calculated.
The spinodal line appears not to be influenced by the
presence of ions at the chosen concentration, while the temperatures of maximum 
density curve displays both a mild shift in temperature and a shape
modification with respect to bulk. Signatures of the presence of a
liquid-liquid critical point are found in the aqueous solution.
By analysing the water-ion radial distribution functions of the aqueous solution 
we observe that upon changing density, structural modifications
appear close to the spinodal. For low temperatures additional 
modifications appear also for densities close to that corresponding to 
a low density configurational energy minimum.
\end{abstract}


%%% the first is computer simulation of liquid structure
%%% the second is metastable phases
%%% the third is structure of liquids 

\maketitle

\section{Introduction}

The anomalous behaviour of water has attracted the attention of
scientists since long time.

Many fascinating features distinguish water from most other
liquids. Among them we point out the presence of a line of temperatures 
of maximum density (TMD) in the $P-T$ plane and the
power law divergences upon supercooling of 
the coefficient of thermal expansion, 
of the specific heat and of the isothermal 
compressibility.~\cite{phystoday,pabloreview,Austin1,Austin2}

Molecular dynamics simulations of water modeled with 
ST2, SPC/E, TIP4P and TIP5P potentials have shown the existence of a second
critical point~\cite{Gene,tanaka,Brovchenko,Poole,Paschek,Vallauri,pasch2} buried 
inside the supercooled region with a coexistence between a low density liquid (LDL) 
and a high density liquid (HDL). The deep supercooled region is not directly accessible to
experiments, nonetheless signatures of the second critical point have been
observed.~\cite{Mishima}
In this picture water anomalies are a direct consequence of the
existence of the second critical point
in the region of low temperatures 
and high pressures, the water spinodal line
is non-reentrant and correspondingly the TMD line bends to avoid 
crossing the spinodal line.

Being water a rather unique solvent, the properties of 
aqueous solutions are also a topic of substantive interest. In particular
thermodynamic and structural properties 
of aqueous solutions of electrolytes are of importance 
because of the ubiquitous presence of these solutions
in biological and geophysical systems. Besides the properties 
of aqueous solutions
upon supercooling are significant for cryopreservation of
biological tissues.~\cite{pablobook,franksbook}

As far as water itself is concerned it is well known that the presence
of solutes affects important properties like melting point, boiling
point and viscosity. 
It is therefore important to understand to what extent thermodynamic
properties of water are affected by the solutes upon supercooling. 
In fact while many studies in literature concentrate on ionic solutions at 
ambient temperature, very little has been done in the supercooled region.
Differential scanning calorimeter measurements have
shown~\cite{Archer} that for low NaCl content, up to about 
1 $mol/kg$, at ambient pressure, the TMD and the 
anomalous behaviour of specific heat upon supercooling are still
present, gradually tending to disappear upon increasing the salt content.

A large number of experiments and simulations focus on the hydration 
structure~\cite{koneshan,leberman,bruni,botti,mancinelli,chowduri,chandra,jardon,hribar,panag,degreve,patra}
at ambient temperature of
ionic solutions, that is to say on the arrangement of water molecules
in proximity to the ions. At low concentrations the presence of the ions 
perturbs the structure of water only locally, while it remains unaltered far
away from the ions.~\cite{hribar}

Traditionally, ions have been divided in 
structure-making (or cosmotrope) and structure-breaking (or chaotrope)
depending on the effect they have on water-water hydrogen bonds.
The first appearance of these concept dates back to the Hofmeister's 
series.~\cite{hofmeister} More than thirty effects connected
to the Hofmeister's series~\cite{kunz} can be found, 
nevertheless its greatest significance 
resides in the different degree of hydration of ionic compounds in aqueous 
solutions. Although some authors~\cite{weck} questioned 
the significance of the structure-making/breaking concept, 
mainly because of the difficulty in quantifying correlated
properties, the majority of studies apply the usual 
definition.~\cite{hribar}  Structure-making ions break  
the hydrogen bonds of surrounding water molecules,
thus allowing a rearrangement of water molecules that results in a ordered 
hydration structure. Structure-breaking ions, 
bend without breaking the hydrogen bonds of nearby water molecules, 
thus disordering the tetrahedrally coordinated network.
So far it is not clear whether and how structure-making/breaking properties of
ions are connected to modifications of thermodynamic properties
of the solvent upon supercooling.  

We present here a molecular dynamics (MD) simulation study of an aqueous 
sodium chloride solution, in the following denoted 
as NaCl(aq), upon supercooling. 
The salt content in the aqueous solution amounts to $c=0.67\, mol/kg$.
Its thermodynamic
and structural properties are compared to a MD simulation of bulk water.  
The isotherms and the isochores are studied
for both systems to investigate 
whether and to what extent the thermodynamic properties of water 
upon supercooling are modified 
in presence of a low concentration of ions. The analysis
of the phase diagrams leads to the determination 
the TMD line and the spinodal line. 

The structural properties are also studied, calculating
the partial radial distribution functions, RDFs,
of water molecules around the ions. The RDFs features are analysed
also taking into account the structure-making properties of the 
$Na^+$ and the structure-breaking properties of the $Cl^-$ ions.

The paper is organized as follows: in Sec.~\ref{model} 
we report the simulation details.
The thermodynamic results are presented in Sec.~\ref{thermo} and discussed
therein. In Sec.~\ref{struct} we present the hydration 
radial distribution functions and 
we discuss their trends as function of density. 
Conclusions are drawn in Sec.~\ref{conclu}.

\section{Model and computer simulation details}\label{model}

Molecular dynamics simulations are performed on two systems,
bulk water and water with $Na^+$ and $Cl^-$ ions.
The systems are simulated with the DL\_POLY package~\cite{dlpoly}
using the TIP4P site model for water molecules.~\cite{tip4p} 

The particles interact via short range L-J 
potentials and long-range coulombic potentials. 
Thus the interaction potential can be written as:

\begin{equation}
U_{ij}(r)=\frac{q_i q_j}{r}+4\epsilon_{ij}
\left[\left(\frac{\sigma_{ij}}
{r}\right)^{12}-\left(\frac{\sigma_{ij}}{r}\right)^{6}\right]
\end{equation}
where ion-water and ion-ion parameters are taken from Koneshan and Rasaiah
calculation for L-J potential,~\cite{koneshan} based upon a re-parametrization of 
Pettitt and Rossky parameters~\cite{pettitt} for the Huggins-Mayer potential. 
The L-J parameter $\epsilon_{ij}$ 
and $\sigma_{ij}$ are obtained using the Lorentz-Berthelot mixing rules
$\epsilon_{ij}=\sqrt{\epsilon_i \epsilon_j}$ and 
$\sigma_{ij}=(\sigma_i +\sigma_j)/2$. The L-J interaction parameters and the charges are
shown in Table \ref{tab:tab1} along with masses and charges of the sites.

\begin{table}[htdp]
\caption{Masses, electric charges and L-J interaction parameters for
TIP4P water sites O, H and X and ions.}
\begin{center}
\begin{ruledtabular}
\begin{tabular}{lccccc}
 & O & H & X & Na$^{+}$ & Cl$^{-}$\\ \hline
m (a.m.u) & 16.000 & 1.008 & -- & 35.453 & 22.990\\
q(e) & -- & +0.52 & -1.04 & +1.00 & -1.00\\
$\epsilon_{O-j}(kJ/mol)$ & 0.65000 & -- & -- & 0.56014 & 1.50575\\
$\sigma_{O-j}(\mbox{\AA})$ & 3.154 & -- & -- & 2.720 & 3.550\\
$\epsilon_{H-j}(kJ/mol)$ & -- & -- & -- & 0.56014 & 1.50575\\
$\sigma_{H-j}(\mbox{\AA})$ & -- & -- & -- &1.310 & 2.140\\
$\epsilon_{\mbox{\scriptsize Na}^+-j}(kJ/mol)$ & 0.56014 & 0.56014 & -- & 0.11913 & 0.35260 \\
$\sigma_{\mbox{\scriptsize Na}^+-j}(\mbox{\AA})$ & 2.720 & 1.310 & -- & 2.443 & 2.796 \\
$\epsilon_{\mbox{\scriptsize Cl}^--j}(kJ/mol)$ &1.50575 &1.50575 & -- & 0.35260 & 0.97906\\
$\sigma_{\mbox{\scriptsize Cl}^--j}(\mbox{\AA})$& 3.550 & 2.140 & -- & 2.796 & 3.487\\
\end{tabular}
\end{ruledtabular}
\end{center}
\label{tab:tab1}
\end{table}
The cut-off radius is set to $9.0\, \mbox{\AA}$ and the long range 
coulombic interactions are treated 
by the Ewald summation method, with the convergence 
parameter $\alpha$~\cite{allen} set to $6.4/L$,
where $L$ is the edge length
of the cubic simulation box in which the particles reside. 

The MD simulations are carried out in the NVT ensemble. 
The temperature is fixed by the use of the Berendsen
thermostat.~\cite{berendsen} 
The integration timestep employed is 1 fs. 

In the case of bulk water 
the system is simulated with 256 water molecules while in the NaCl(aq) 
system we have 250 water molecules plus 3 sodium ions and 3 chloride ions. 
The corresponding concentration expressed as moles of solute per mass 
of solvent (molality) is $c=0.67\, mol/kg$.

We span densities in the range 
$0.80\, g/cm^3 \leq \rho \leq 1.1\,g/cm^3$ for the NaCl(aq) and
$0.80\, g/cm^3 \leq \rho \leq 1.05\, g/cm^3$ for bulk water.
The analysed densities and the corresponding boxlenghts are reported
in Table \ref{tab:tab2} for both systems. 
In the starting  configurations particles lie on a face centered cubic lattice 
with random orientation  of water molecules. The crystal is melted at 500 K and temperature is then
progressively decreased during the equilibration runs. 
The longest equilibrations have been run for 10 ns. 
For both systems we examine 
temperatures in the range $210\, \mbox{K} \leq T \leq 500 \, \mbox{K}$.
Some of the thermodynamic points for the bulk system are taken from a previous 
work of our group.~\cite{gallorovere}

\begin{table}[htbp]
\caption{Densities and boxlenghts for NaCl(aq) and bulk
  water.}
\begin{center}
\begin{ruledtabular}
\begin{tabular}{ccc}
$\rho\, (g/cm^3)$ & NaCl(aq) $L_{box}$(\AA) & Bulk $L_{box}$(\AA) \\ \hline
1.1 & 19.189 & --\\
1.05 & 19.489 & 19.388\\
1.025 & 19.641& -- \\
1.0 & -- & 19.710\\
0.98 & 19.943 & 19.839\\
0.95 & 20.151 & 20.046 \\ 
0.90 & 20.517 & 20.410\\ 
0.87 & 20.750 & 20.650\\ 
0.85 & 20.911 & 20.811\\ 
0.83 & -- & 20.977\\
0.80 & 21.338 & 21.236\\
\end{tabular} 
\end{ruledtabular}
\label{tab:tab2}
\end{center}
\end{table}

\begin{figure}[b]
\centerline{\psfig{file=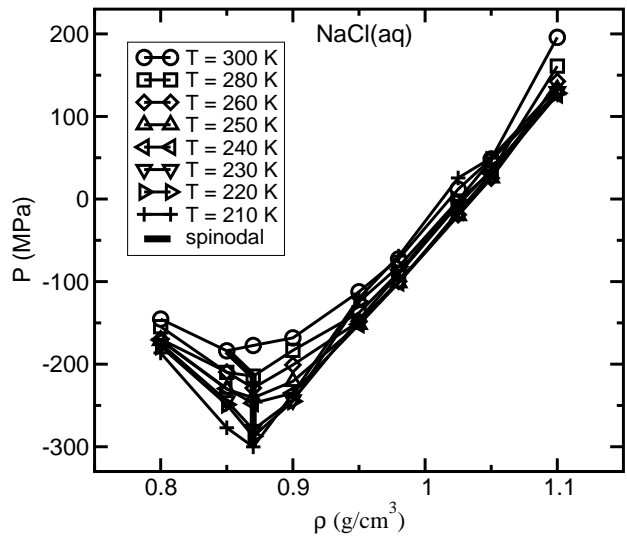,width=0.5\textwidth}}
\caption{Spinodal line and isotherms of the NaCl(aq) solution in the $P-\rho$ plane.}
\label{fig:1}
\end{figure}

\section{Thermodynamic Results}\label{thermo}

We investigate how the thermodynamic behaviour
of water upon supercooling changes in the presence 
of a low concentration of $Na^+$ and $Cl^-$ ions with 
respect to the bulk by looking at isotherms and isochores
obtained from simulated state points. 

From the isotherms the limit of mechanical stability can be determined
in the framework of mean field theories from the divergence of the
isothermal compressibility. The singularity points where 
\begin{equation}
\left( \frac{\partial P}{\partial \rho} \right)_T=0
\label{eq:spinod}
\end{equation}
define the spinodal line. They
can be obtained searching for the minima of the $P_T(\rho)$ curves.

\begin{figure}[b]
\centerline{\psfig{file=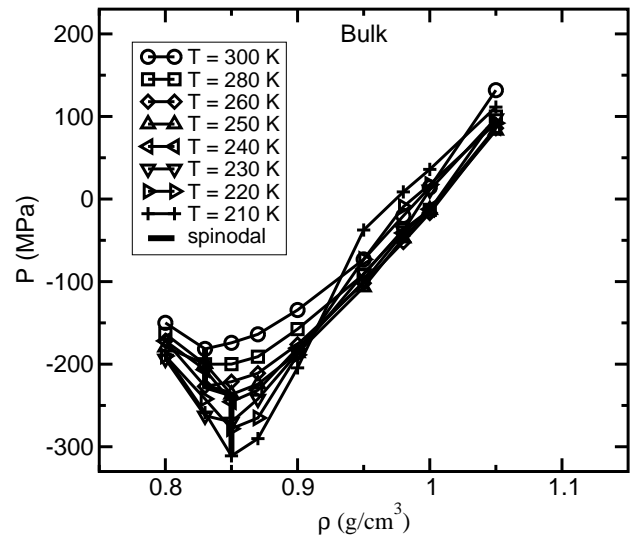,width=0.5\textwidth}}
\caption{Spinodal line and isotherms of bulk water in the $P-\rho$ plane.}
\label{fig:2}
\end{figure}

In Fig.~\ref{fig:1} we show the isotherms of the NaCl(aq) system obtained upon 
cooling for densities ranging from  $\rho=0.80\, g/cm^3$
to  $\rho=1.1\, g/cm^3$ and temperatures from $T=300$~K to
$T=210$~K. The spinodal line has been drawn joining the minima 
of the isotherms. The isotherms from $T = 500$~K to $T=350$~K do not show 
a minimum and they have been not reported. The $T = 300$~K 
isotherm is the only one to have the minimum in correspondence of the
density $\rho=0.85\, g/cm^3$, 
all the other lower temperature isotherms show a minimum corresponding to the density 
$\rho=0.87\, g/cm^3$. 

We report in Fig.~\ref{fig:2} the isotherms of bulk
water in the $P-\rho$ plane. We plot the bulk isotherms for densities ranging from 
$\rho=0.80\, g/cm^3$ to $\rho=1.05 \, g/cm^3$ and temperatures 
ranging from $T=300$~K to $T=210$~K. Analogous to 
what happens for the solution the highest temperatures 
isotherms, $T=500$~K, $T=400$~K and $T=350$~K do not exhibit
minima (not shown).
The spinodal line for bulk water, drawn 
joining the isotherms minima, shifts from $\rho=0.83\, g/cm^3$, for $T$ 
ranging from $300$~K to $260$~K,
to $\rho=0.85 \, g/cm^3$, for the lowest temperatures.

\begin{figure}[h]
\centerline{\psfig{file=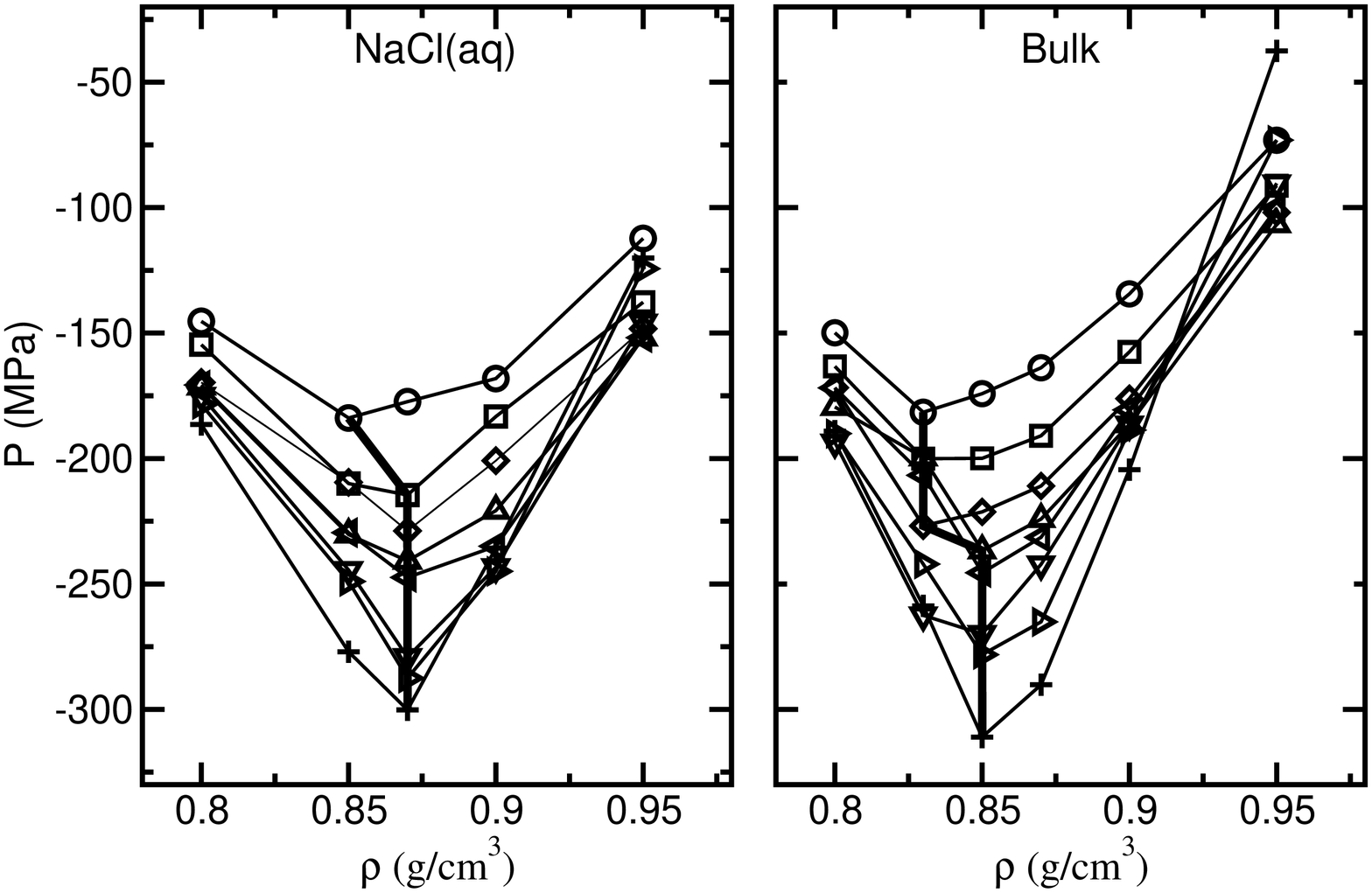,width=0.5\textwidth}}
\caption{Blow up and comparison of the NaCl(aq) and bulk spinodal line and isotherms in the $P-\rho$ plane,
in the density range $0.80\, g/cm^3\leq \rho \leq 0.95 \, g/cm^3$ . Symbols are as in Fig.~\ref{fig:1} and Fig.~\ref{fig:2}.}
\label{fig:3}
\end{figure}

The comparison of the region of the $P-\rho$ plane close to the spinodal line 
of the two systems is reported in Fig.~\ref{fig:3}. The density range is 
here $0.80\, g/cm^3\leq \rho \leq 0.95 \, g/cm^3$. We observe that
the NaCl(aq) isotherms minima are shifted to higher densities with respect to the bulk, moreover the curves envelope is
globally shifted to lower pressures.
We note that the presence of the ions 
also changes the shape of the curves. The isotherms of 
the NaCl(aq) system appear in fact closer to each other than in 
the bulk phase.

\begin{figure}[h]
\centerline{\psfig{file=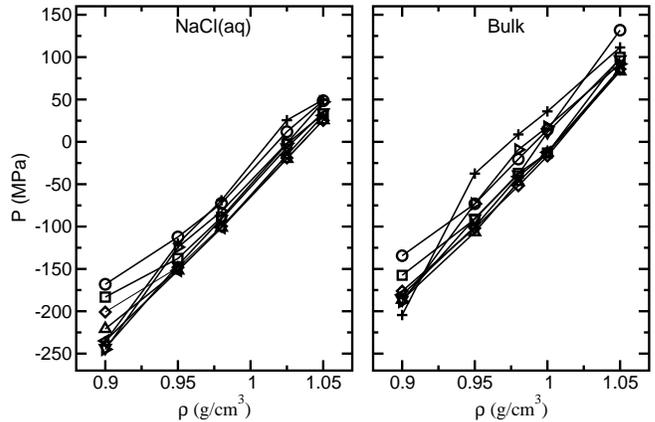,width=0.5\textwidth}}
\caption{Blow up and comparison of the NaCl(aq) and bulk isotherms in the $P-\rho$ plane,
in the density range $0.90\, g/cm^3\leq \rho \leq 1.05 \, g/cm^3$. Symbols are as in Fig.~\ref{fig:1} and Fig.~\ref{fig:2}.}
\label{fig:4}
\end{figure}

In Fig.~\ref{fig:4} we present 
a zoom of the region of the
$P-\rho$ phase diagram in the density range
$0.90\, g/cm^3\leq \rho \leq 1.05 \, g/cm^3$, in order to compare the 
isotherms of the
two systems at high densities. In both systems 
the $T=220$~K and the $T=210$~K curves 
cross the higher temperature isotherms and change the sign of
their curvature. These effects are particularly evident 
for the bulk $T=220$~K and $T=210$~K isotherms and 
the NaCl(aq) $T=220$~K isotherm. In bulk water 
these changes of curvatures have been 
already observed~\cite{Gene,spinodal1} and related to the existence
of a second critical point. In fact they are 
signatures of the approach of the system to the coexistence 
between low an high density liquid phases. Therefore we hypothesize
also for the  NaCl(aq) the existence of a second critical point.
For the change of curvature a shift toward higher 
densities can be noticed in the aqueous solution. 
This observation leads to infer that also the second 
critical point might consequently be shifted
to a higher density with respect to the position in the bulk.

\begin{figure}[t]
\centerline{\psfig{file=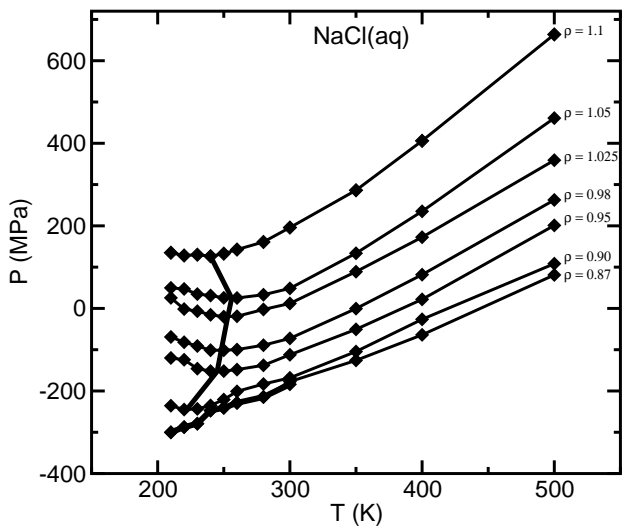,width=0.5\textwidth}}
\caption{Spinodal line, TMD line and isochores of the NaCl(aq) solution in the $P-T$ plane.}
\label{fig:5}
\end{figure}

\begin{figure}[t]
\centerline{\psfig{file=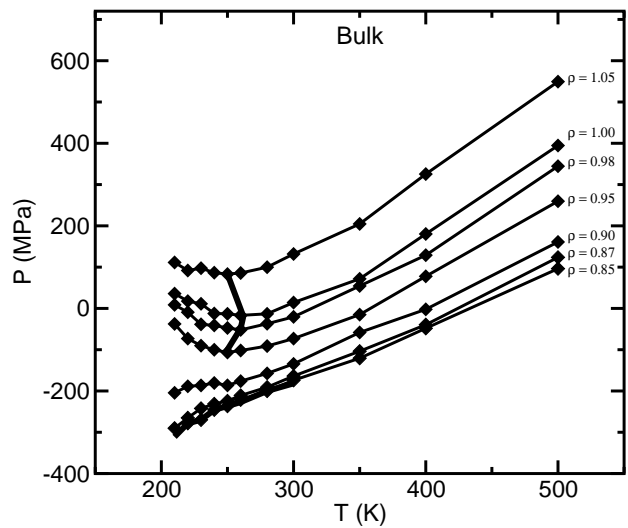,width=0.5\textwidth}}
\caption{Spinodal line, TMD line and isochores of bulk water in the $P-T$ plane.}
\label{fig:6}
\end{figure}

We now consider the isochores of the NaCl(aq) and the pure water systems
reported in Fig.~\ref{fig:5} and Fig.~\ref{fig:6} respectively.
The behaviour of the TMD line can be extracted from the analysis
of the isochores plane. In fact along the TMD line 
the coefficient of thermal expansion
\begin{equation}
\alpha_P=\frac{1}{V} \left( \frac{\partial V}{\partial T} \right)_P
\end{equation} 
goes to zero.
The thermal pressure coefficient 
\begin{equation}
\gamma_V= \left( \frac{\partial P}{\partial T} \right)_V
\end{equation} 
is connected to the coefficient of thermal expansion by the following
equation
\begin{equation}
\gamma_V=\frac{\alpha_P}{K_T}\, .
\end{equation}
where $K_T$ is the isothermal compressibility.
Therefore the TMD points lie on the line connecting 
the minima of the isochores. For both systems the minima have been determined
fitting the isochores with a fourth order polynomial function.

For the aqueous solution the isochores in Fig.~\ref{fig:5} are
shown from $\rho=1.1 \, g/cm^3$ to  $\rho=0.87 \, g/cm^3$ 
in the temperature ranges from $T=500$~K to $T=210$~K.
The $\rho=0.87 \, g/cm^3$ isochore almost coincides with the spinodal line
(see Fig.~\ref{fig:1}) also reported in Fig.~\ref{fig:5}.
In Fig.~\ref{fig:6} the isochores of the bulk water are reported from
$\rho=1.05 \, g/cm^3$ to $\rho=0.85 \, g/cm^3$, in the same temperature range
as above. Here the $\rho=0.85 \, g/cm^3$ isochore almost
coincides with the water spinodal shown in the same figure.
For both systems the isochores  with densities lower than that of the
the spinodal lines are not reported since
they lie beyond the mechanical stability limit of the system.

For bulk water we observe a spinodal extending to the region of negative pressures and non 
reentrant down to the lowest temperature investigated. This behaviour has been already observed in
literature for different potentials.~\cite{spinodal1,Essmann,Poole,tanaka,hpss,netz,Mossa,minozzi}
Also the aqueous solution displays a non reentrant spinodal line
qualitatively similar to the bulk.

In the case of the aqueous solution all the isochores 
above $\rho=0.87 \, g/cm^3$ exhibit a clear minimum.
This result agrees with the
experimental work by Archer and Carter~\cite{Archer} 
in which they find a TMD line for the solution at the
concentration studied here.  
From the minima, via a fitting procedure, we determine the exact
location of the TMD.
We observe that the TMD line, reported in Fig.~\ref{fig:5}, 
reverts its path avoiding the spinodal. 

In the case of pure water (Fig.~\ref{fig:6}) only for the first
four isochores a minimum is found in the range of
temperatures explored. 
The shape of the TMD line is in agreement with the results that can
be found in literature.~\cite{Poole,spinodal1}

The comparison of the isochores planes of 
NaCl(aq) (Fig.~\ref{fig:5}) and bulk water (Fig.~\ref{fig:6}) 
reveals that in the $P-T$  plane the isochores of NaCl(aq) 
lie at lower pressures than the corresponding bulk curves. 
Besides, the concavity of the isochores in proximity 
to the minima is deeper in bulk water than in NaCl(aq). 
This is mirrored by the different pattern followed 
by the TMD line in the two systems.

The location of the second critical point for bulk TIP4P water
is estimated to be at negative pressure by Tanaka~\cite{tanaka},
while it is found at positive pressures for ST2~\cite{Poole} 
TIP5P~\cite{Paschek} and TIP4P-ew~\cite{pasch2}. 
From the behaviour of our iscohores, TMD and spinodal for both the bulk 
water and the aqueous solution it is actually not easy 
to establish whether it falls at positive or negative pressures.

\begin{figure}[h]
\centerline{\psfig{file=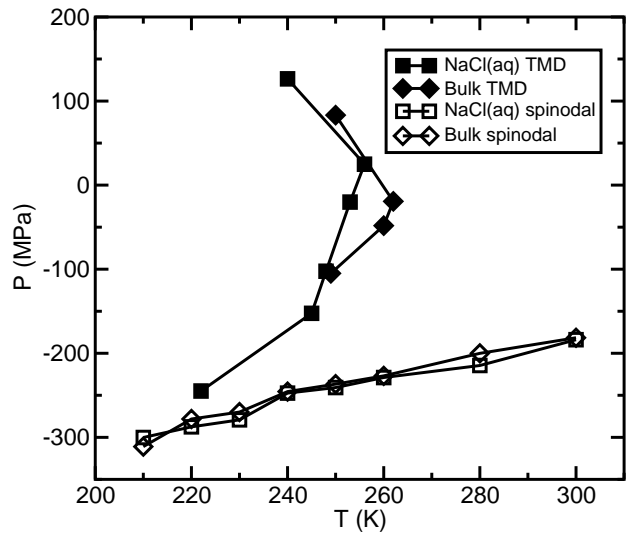,width=0.5\textwidth}}
\caption{TMD and spinodal lines for NaCl(aq) and bulk water in the $P-T$ plane.}
\label{fig:7}
\end{figure}

In Fig.~\ref{fig:7} the TMD line of NaCl(aq) and bulk water 
are reported together with 
the spinodal lines of the two systems. Comparing the TMD lines 
we notice that, apart for the mild 
temperature shift due to the colligative properties of
the solutions, the TMD lines appear different in shape. 
In fact, spanning the same range of densities, the 
NaCl(aq) TMD extends in a region of pressures wider than that 
of the bulk. Therefore, the effect of the ions 
upon supercooling is not limited to give
a rigid translation of the phase diagram. 

While the presence of the ions seems to induce important changes in an 
anomalous thermodynamic property of water, the TMD, 
the limit of mechanical stability appears not affected
by the ions. In fact from Fig.~\ref{fig:7} 
we can see that the spinodal lines of the two systems
are almost indistinguishable from each other.

\begin{figure}[h]
\centerline{\psfig{file=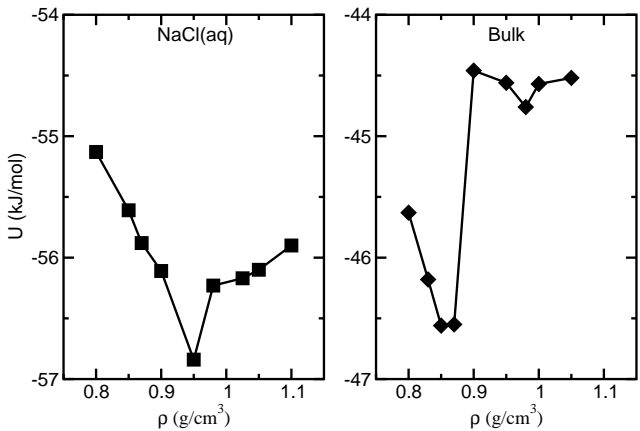,width=0.5\textwidth}}
\caption{Comparison of the configurational 
energy per molecule, at $T=220$~K, as a function of the density.}
\label{fig:8}
\end{figure}

In Fig.~\ref{fig:8} we present the configurational 
(or potential) energy (per molecule) 
of the NaCl(aq) and bulk systems, at $T=220$~K, as a function of
density. In the NaCl(aq)
U presents a minimum in correspondence of $\rho=0.95 \, g/cm^3$ while in the bulk 
the minimum is reached for $\rho=0.86 \, g/cm^3$. These energy extrema can
be related to the existence of LDL and HDL phases.
It has been shown by Kumar \emph{et al.}~\cite{Kumar} that 
two minima are present for TIP5P water confined between hydrophobic plates
for $\rho=0.88 \, g/cm^3$ and for $\rho=1.39 \, g/cm^3$. 
The presence of this minimum in our system 
supports the existence of a  liquid-liquid (LL)
coexistence also in the aqueous solution.

\section{Structural Results}\label{struct}

We discuss in this section the radial distribution functions (RDFs) of the 
aqueous solution, $g(r)$,  computed averaging the 
equilibrated configurations. 

In the NaCl(aq) system 
at the salt content investigated we do not observe significant changes 
in water-water RDFs  with respect to the bulk system.
Minor differences occur only in the secondary peaks, 
for low densities and in the deep supercooled region (not shown).
It has been shown in a number of
papers~\cite{leberman,bruni,botti,mancinelli} 
that the effect of the ions on water-water structure
is similar to the one obtained applying an external pressure
and that it becomes more evident upon increasing the salt content. 
The salt concentration in our system, $c=0.67\, mol/kg$, 
is probably too low  
to find clear evidence of this effect. Besides both sodium and
chloride ions can be found in the central region of the Hofmeister
series so that not very marked effect on water-water structure are expected.

We present in the following the analysis of the water-ion structure 
comparing RDFs along the isotherms $T=300$~K and $T=220$~K for
densities from $\rho=1.1\, g/cm^3$ to $\rho=0.80\, g/cm^3$.
We note that for the highest density at $T=300$~K the RDFs show
different trends with respect to all other densities investigated.
These trends likely anticipate a different behaviour of the system 
at very high densities (not investigated in the present paper).

\begin{figure}[htbp]
\centerline{\psfig{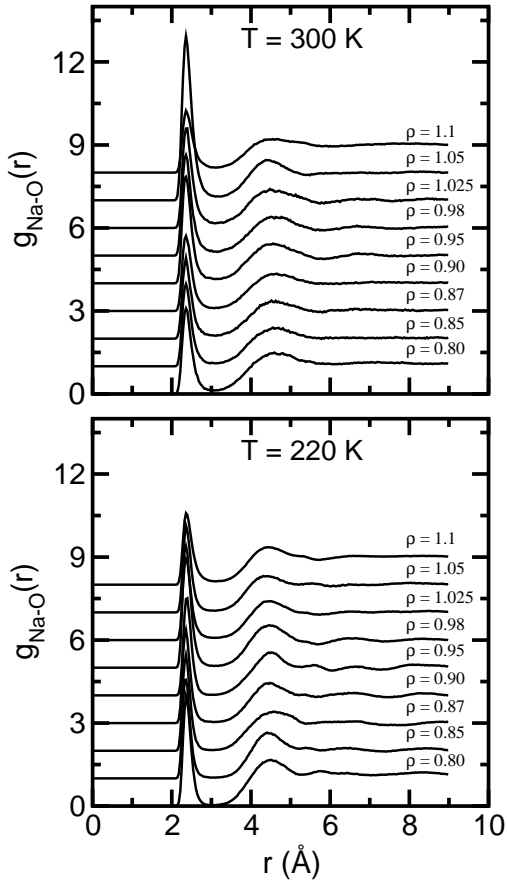}}
\caption{Na--O radial distribution functions at $T=300$~K (top panel) and $T=220$~K (bottom panel). 
The curves are shifted by a unit step on the y-axis for the sake of clarity.}
\label{fig:9}
\end{figure}

\begin{figure}[htbp]
\centerline{\psfig{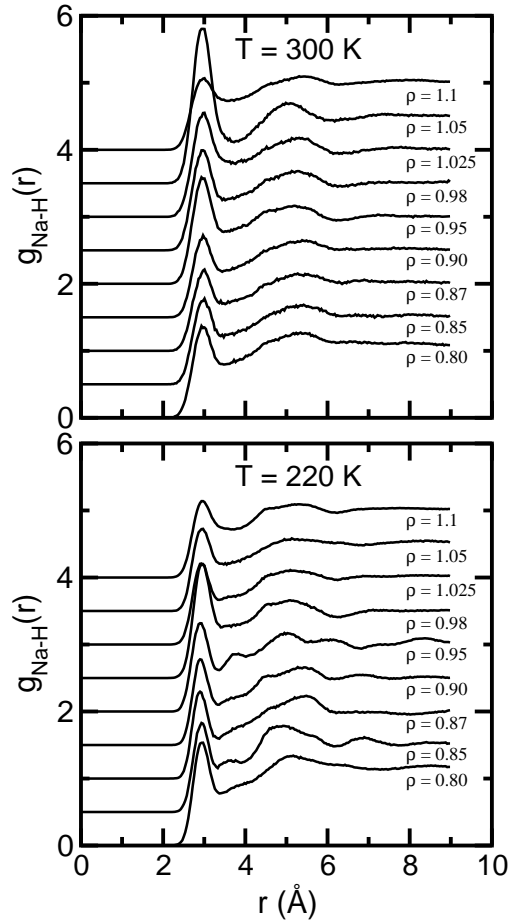}}
\caption{Na--H radial distribution functions at $T=300$~K (top panel) and $T=220$~K (bottom panel).
The curves are shifted by a step of 0.5 on the y-axis for the sake of clarity.}
\label{fig:10}
\end{figure}

In Fig.~\ref{fig:9} and Fig.~\ref{fig:10} we show the sodium-oxygen and
the sodium-hydrogen RDFs.

We observe that for both temperatures the first and the second
coordination shells of the $g_{Na-O}(r)$ are well defined  
showing the tendency of water to form stable hydration shells
around the sodium ion, usually classified as a structure-making.
For the $g_{Na-H}(r)$ the first shell is also well defined
while the second shell is broader.

\begin{figure}[ht]
\centerline{\psfig{file=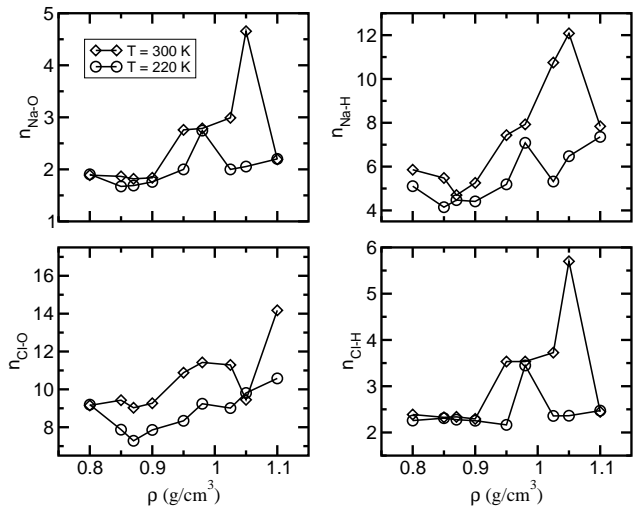,width=0.5\textwidth}}
\caption{Coordination numbers relative to the first shell for the Na-O, Na-H, Cl-O and Cl-H 
pairs. }
\label{fig:10bis}
\end{figure}

The position of the first peak of the $g_{Na-H}(r)$ is shifted 
to a higher distance with respect
to the position of the first peak of the $g_{Na-O}(r)$. The difference
between the two peaks 
is circa $0.6\,\mbox{\AA}$ indicating that the two
hydrogens of the molecule lie symmetrically around the sodium ion.
This result is confirmed by the coordination numbers of the first 
shell which we report in Fig.~\ref{fig:10bis} together with the
coordination numbers of the Cl--O and Cl--H.

We consider now more in detail the behaviour of the  $g_{Na-O}(r)$
as function of the density.

At $T=300$~K starting from $\rho=1.05\, g/cm^3$  
the peaks heights decrease significantly
until they reach a  
minimum at the density of the spinodal line,
$\rho=0.87\, g/cm^3$. For densities lower than this value
down to $\rho=0.80\, g/cm^3$  the peaks heights slightly increase
again. On going from high density down to the spinodal density 
along the isotherm the $Na^+$ ions partially lose 
their structure-making ability due to the
approach to the instability region, where the onset of large
fluctuations and a tendency to demixing could appear.

At $T=220$~K the two peaks slightly sharpen as a consequence of 
decreasing temperature. Starting from  $\rho=1.05\, g/cm^3$
the height of the first peak of the $g_{Na-O}(r)$ 
now shows a maximum in proximity of the energy minimum density, then it
decreases showing a minimum on approaching the spinodal density.
We observe close to the second peak 
the appearance of secondary shells 
especially for densities close to the spinodal and to 
the energy minimum density, e.g. 
a possible region of LL coexistence.
This indicates that the second shell of
hydration is less defined.

The $g_{Na-H}(r)$, shown in Fig.~\ref{fig:10}
displays an analogous trend 
as a function of density both for $T=300$~K and $T=220$~K
with respect to that 
observed in Fig.~\ref{fig:9} for the sodium-oxygen pair.
In particular we observe also here the appearance of secondary shells.

\begin{figure}[ht]
\centerline{\psfig{file=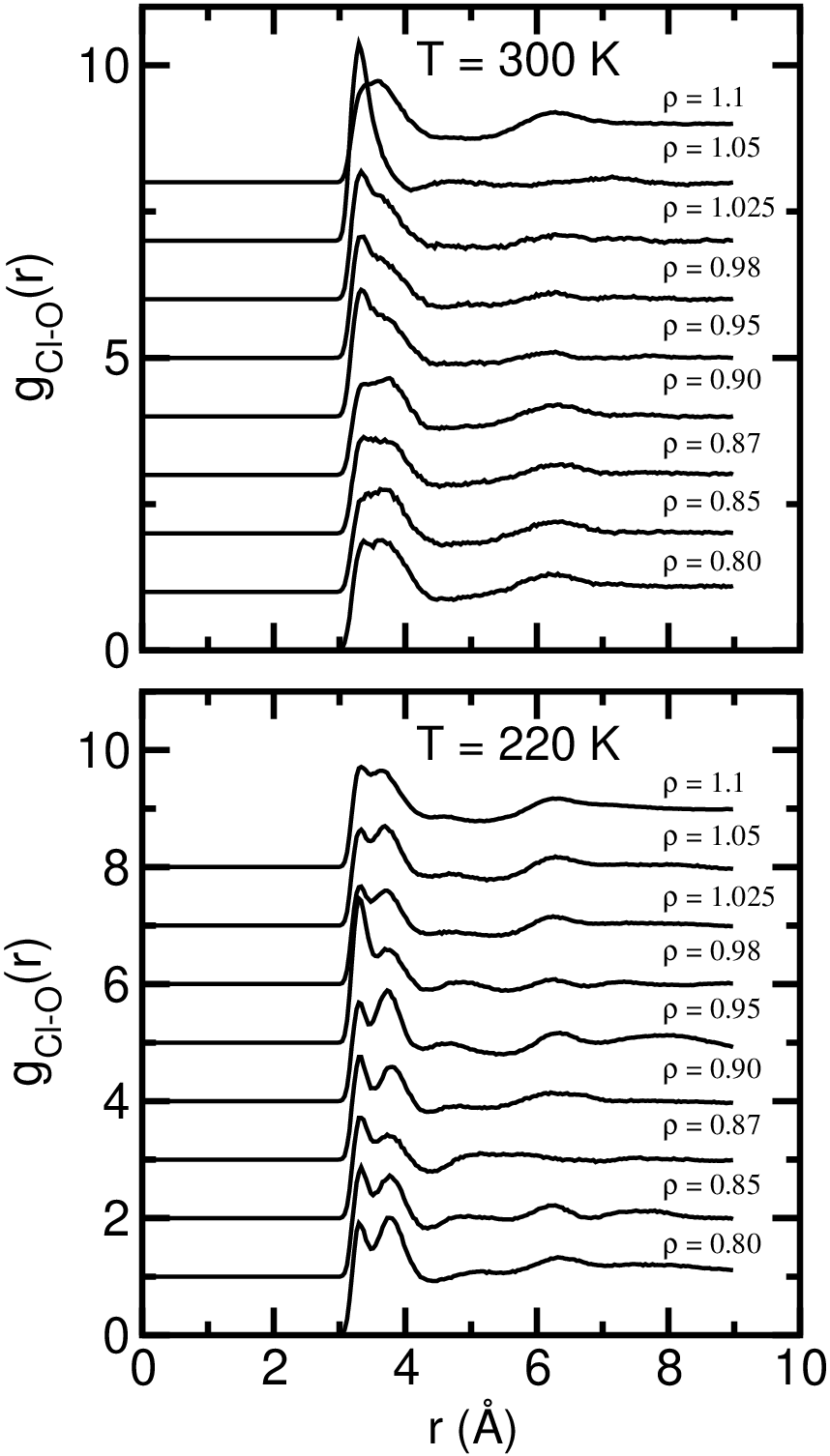,width=0.37\textwidth}}
\caption{Cl--O radial distribution functions at $T=300$~K (top panel) and $T=220$~K (bottom panel)
The curves are shifted by a unit step on the y-axis for the sake of clarity.}
\label{fig:11}
\end{figure}

\begin{figure}[ht]
\centerline{\psfig{file=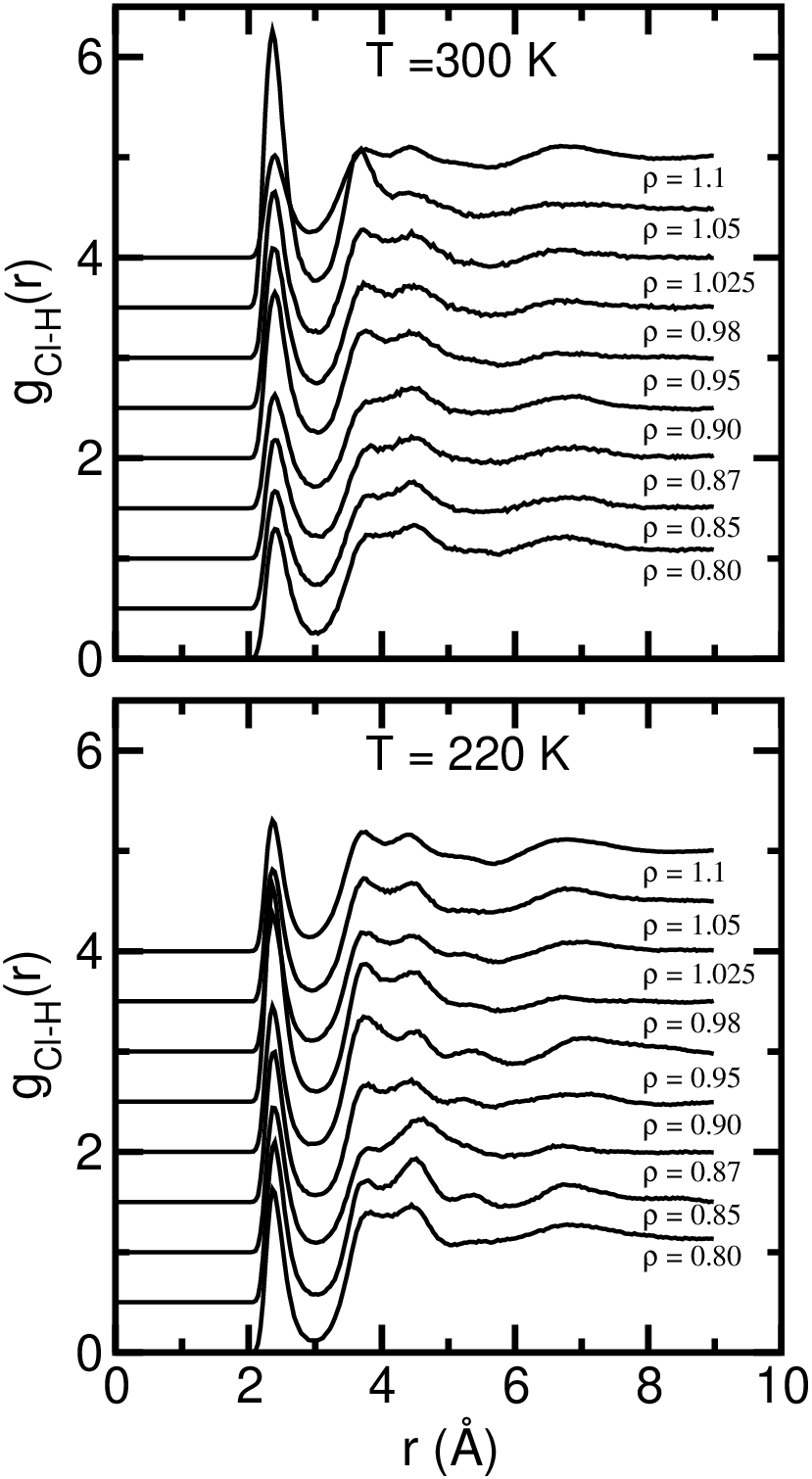,width=0.37\textwidth}}
\caption{Cl--H radial distribution functions at $T=300$~K (top panel) and $T=220$~K (bottom panel).
The curves are shifted by a step of 0.5 on the y-axis for the sake of clarity.}
\label{fig:12}
\end{figure}

In Fig.~\ref{fig:11} and Fig.~\ref{fig:12}  we show the $g_{Cl-O}(r)$ and the
 $g_{Cl-H}(r)$.

For the $g_{Cl-O}(r)$ we see
a broad and structured first shell and a not well defined second shell.
For the $g_{Cl-H}(r)$ we observe a well defined first peak 
and a broad and structured second shell.
At variance with the  $g_{Na-O}(r)$ and the
$g_{Na-H}(r)$  the first and the second shell are not always 
well defined for these RDFs 
because of the structure breaking property of the $Cl^-$ ion.

By comparing the positions of the first peaks of 
the $g_{Cl-O}(r)$ and the $g_{Cl-H}(r)$
we note an approximated charge ordering around the negative
chloride ion. The partially positive charged hydrogens move closer
to the ions and the first shell of the oxygens is placed around
the first minimum of the $g_{Cl-H}(r)$.
 
We now analyse the behaviour of the $Cl^-$ RDFs as a function of densities
for the two temperatures investigated.

For the $g_{Cl-O}(r)$
at $T=300$~K the first shell of oxygens is well defined only at 
$\rho=1.05\, g/cm^3$. 
Below $\rho=1.05\, g/cm^3$ a shoulder appears on the right side of
the first peak. The peaks heights of the $g_{Cl-O}(r)$ 
show a trend similar to that of the $g_{Na-O}(r)$ while 
their shape is more strongly dependent on the density.
On approaching the spinodal the
first peak merges with the shoulder resulting in a broad structure.
The structure-breaking role of the $Cl^-$ ions seems to be reinforced
close to the instability limit.

At $T=220$~K the shoulder close to the first peak evolves in a
second peak indicating a less stable configuration of the chloride
hydration with respect to sodium. 
In fact the first hydration shell is characterized 
by the existence of two possible equilibrium positions
albeit close to each other for the chloride-oxygen pair.
In this first shell oxygens of two kinds are present, those
tied to the hydrogens 
of the first shell of the $g_{Cl-H}$ and those
tied to hydrogens belonging to the  second shell of the $g_{Cl-H}$
that in fact also shows double peak 
structure (see Fig.\ref{fig:12}).
For this temperature a structuring of the curves also appears
for densities close to the energy minimum density 
and to the spinodal density. 

For the $g_{Cl-H}(r)$, Fig.~\ref{fig:12}, the
trend of the peaks intensities as a function of density
are similar to those in Fig.~\ref{fig:11}
but the RDFs appear much more structured with a second and third shell
evident at all densities. The second shell is modulated in a double
peak structure for both temperatures.
We also note that
for both temperatures the first shell of hydrogens of the $g_{Cl-H}(r)$
is well defined at all densities, at variance with the
second peak. It shows that hydrogens penetrate inside 
the shell of the oxygens for the attraction of the ions. 

The broadening and the modulation of the oxygen first peaks
and the hydrogen second peaks in the the  $g_{Cl-O}(r)$
and the  $g_{Cl-H}(r)$ respectively
seem to be related to a competition
between the formation of the hydrogen bonds and the repulsion/attraction
of the chloride ion with respect to oxygen/hydrogen atoms.

We have seen that the $Na^+$ ions are able to form a sufficiently 
rigid shell of oxygens, see the first peaks in  Fig.~\ref{fig:9}, 
breaking the hydrogen bonds when
necessary while the $Cl^-$ ions can attract the hydrogens 
in a first shell and repel the oxygens, but they have
little effect on the hydrogen bonding. 
Coordination numbers confirm this picture, see Fig.~\ref{fig:10bis}.
They show in fact no correlation in trend between the first
shell of hydrogen and the first shell of oxygens 
around the chloride ion as a function of 
density.

For each of the four hydration RDFs the peaks positions are compatible with 
what found in literature.~\cite{koneshan,chowduri,chandra,jardon}

The ion-ion RDFs (not shown) have a low statistic due to the low concentration.
The positions of the peaks indicate the presence of clusters of
positive and negative ions
in agreement with what reported in several 
papers,~\cite{koneshan,panag,degreve,georgalis}
both theoretical and experimental.

\section{Conclusions}\label{conclu}

We performed molecular dynamics simulations of a sodium chloride 
aqueous solution with concentration $c=0.67\, mol/kg$ and of bulk water 
upon supercooling  at different densities. 
We used TIP4P water model both for the aqueous solution and for the bulk.

We studied how the thermodynamic behaviour 
of bulk water is modified by the presence of ions 
and in particular the possible modification of the TMD and of the 
spinodal line. 
We found that while the presence of ions influences the shape and the
position of the TMD, it leaves substantially unaltered the spinodal
line with respect to the bulk. Besides
the TMD curve appears broadened and it extends to much lower
temperatures than bulk TMD.
In a previous study of TIP4P water under hydrophobic
confinement~\cite{gallorovere} 
it was found that both the spinodal line and the TMD are shifted toward
higher pressures. In the case of the aqueous solution 
studied in the present work
the TMD line shifts also to lower temperatures
and correspondingly its curvature broadens.
In addition, the presence of the ions moves the isotherms and the isochores 
toward lower pressures and it slightly modifies their shape, especially in 
proximity to the minima of those
curves. Therefore, although the limit of mechanical stability of the system 
remains unaltered in
the solution, the effect of ions is not limited to a trivial pressure shift. 
The modification of the 
isotherms and the isochores curves and especially of the TMD shows that in 
the supercooled regime 
the ions affect the thermodynamics of the system even at this low 
concentration. 

In our aqueous solution we observed at higher densities a  
crossing of the two lowest temperatures isotherms with the higher 
temperatures ones. It has been shown that the 
change of sign of the curvature of isotherms in this density region is
connected to the existence of a second critical 
point in bulk water.~\cite{Gene,spinodal1} We 
can consequently hypothesize the existence of a LL critical 
point also in the solution probably shifted to higher densities. 
The TMD shape is related to the position of the LL critical point hence
our results indicate that the presence of ions
stabilizes the high density liquid phase driving down in temperature
the LL line. 
Interestingly two recent experimental papers by 
O. Mishima~\cite{mishi1,mishi2} on LiCl-water solutions
seem to indicate a suppression of the transition to the low density 
amorphous-like (LDA-like) water. 
This appears consistent with our TMD line shift.
Besides, according to J. Holzmann et al. the shift to low temperatures 
of the TMD can be framed in a picture in which the free water,
that is the water far from ions, becomes upon supercooling 
more mobile and pressurized.~\cite{pasch}

The spinodal line, related to low density liquid 
properties, does not show any substantial modification 
with respect to the bulk at least when a small amount of ions
is present.

The shape of the curve describing the configurational 
energy as a function of the 
density (see Fig. \ref{fig:8}) for the NaCl(aq) closely resembles 
what found by Kumar \emph{et al.}~\cite{Kumar} 
for confined water in the low density region zone explored in this work. 
Importantly we observe the presence of a
minimum analogous to the LL low density coexistence minimum 
observed in bulk water.

The structural analysis of the NaCl(aq), 
shown here for $T=300$~K and at $T=220$~K, 
revealed that
the four hydration radial distribution functions $g_{Na-O}(r)$, $g_{Na-H}(r)$,
$g_{Cl-O}(r)$ and $g_{Cl-H}(r)$  display minima 
in peaks intensities in proximity to the 
density corresponding to the spinodal line at both high and low
temperatures and the appearance of several 
secondary peaks for densities close to the spinodal line and to the 
configurational energy minimum, at low temperatures. Therefore we found that 
the spinodal line and the energy minimum, low density LL
coexistence, affect significantly the 
hydration structure. 

As regards structure-making/breaking effects, 
the traditional classification of $Na^+$ ion as 
structure-making and $Cl^-$ ion 
as structure-breaking~\cite{jardon,hribar,gurney,bockris} 
appears confirmed in this work. Sodium ions in fact tend to break 
water hydrogen bonds 
favouring a rearrangement of the equilibrium positions of the molecules 
around the ion
that results in the formation of definite hydration shells 
(especially the first one). On the other
hand, the chloride having a greater ionic radius and thus a minor 
charge density, is less effective
in breaking the hydrogen bonds.  Therefore its coordination shells 
are less defined and more disordered as also indicated by
the presence of secondary peaks in the shell structures.

We have seen that all peaks intensities of our hydration RDFs 
follow a similar pattern as a function of the density. 
The presence of minima in the peaks heights is probably 
connected to the approach
of the system to the spinodal line.

On going from high density toward the spinodal density 
$Na^+$ ions partially loose their structure making ability
while the structure breaking properties of the $Cl^-$ ions seems
to be reinforced.

Several extensions of this study can be proposed. 
In particular the analysis of higher densities
and concentrations would be of the uttermost 
interest in order to investigate how thermodynamic and
structural properties change varying those variables. 
Indeed varying either the 
concentration or the chemical species of the solutes provides a tool to 
prepare a system with suitable properties, for example for cryopreservation.

\section{Acknowledgments} We thank S. Buldyrev and V. Molinero for
stimulating discussions.

\end{document}